\documentstyle[preprint,pre,aps,eqsecnum]{revtex}
\tightenlines
\begin{document}


\title{Phase-field model for Hele--Shaw flows\\
with arbitrary viscosity contrast. II. Numerical study}

\author{
R. Folch, J. Casademunt, A. Hern\'andez--Machado}
\address{
Departament d'Estructura i Constituents de la Mat\`eria\\
Universitat de Barcelona,
Av. Diagonal, 647, E-08028-Barcelona, Spain
}
\author{L. Ram\'{\i}rez--Piscina}
\address{
Departament de F\'{\i}sica Aplicada\\
Universitat Polit\`ecnica de Catalunya,
Av. Dr. Mara\~n\'on, 50, E-08028-Barcelona, Spain
}

\maketitle

\begin{abstract}

We implement a phase-field simulation of the dynamics of two fluids 
with arbitrary viscosity contrast in a rectangular Hele-Shaw cell.
We demonstrate the use of this technique in different situations including
the linear regime, the stationary Saffman-Taylor fingers and the 
multifinger competition dynamics, for different viscosity contrasts.
The method is quantitatively tested against analytical predictions and
other numerical results. A detailed analysis of convergence to the sharp 
interface limit is performed for the linear dispersion results.
We show that the method may be a useful alternative to  
more traditional methods. 


\vskip 5mm
\centerline{Copyright 1999 by The American Physical Society}

\end{abstract}

\newpage
\section{introduction}

In the previous paper \cite{teoric} we presented a phase-field model for the 
Saffman--Taylor problem with arbitrary viscosity contrast. 
The problem consists in determining the time evolution of the
interface between two inmiscible (viscous) fluids within a Hele--Shaw cell
\cite{st}.
This can be done analytically only in very restricted situations,
so numerical
work is usually required.
As other free boundary problems (bulk problems with boundary conditions
on a moving boundary), Hele--Shaw flows have traditionally been
formulated by projecting the bulk dynamics onto the interface
using boundary-integral methods which lead to integro-differential equations
\cite{aref1,aref2,aurora,jaume,vinals,shelley}. An alternative 
approach, namely the so-called
phase-field model, has also been used to study solidification and
related problems
\cite{langer,kobayashi,mcfadden,karma,grant,heine,mozos,ricard}. This approach 
describes the system in terms of a set of partial differential equations 
avoiding the treatment of the interface as a moving boundary. To do so, 
an additional field is introduced which locates the interface at a region 
of thickness $\epsilon$. The equations are then required to yield the 
original moving-boundary problem in the sharp interface limit 
$\epsilon \rightarrow 0$.
For Hele--Shaw flows, such a phase-field model was introduced in Ref. 
\cite{teoric}. In this case two independent
small parameters ($\epsilon$ and $\tilde{\epsilon}$) were actually introduced,
over which three
distinct conditions control the convergence to the 
sharp-interface limit $\epsilon,\tilde{\epsilon} \rightarrow 0$.
In the paper we proved that model to yield the right 
sharp-interface equations in this limit and obtained a 
`thin-interface' model, which consists of a set of effective sharp-interface
equations which keep finite-$\epsilon$ and -$\tilde{\epsilon}$
effects up to first order. 
This provides appropriate criteria
to choose the computational parameters for numerical simulations, 
and, for the linear regime, it enables
to explicitly compute deviations from the Hele--Shaw growth rates.

The purpose of the present paper is to numerically check the phase-field model, 
and eventually the corresponding thin-interface approximation,
against known
sharp-interface solutions, and to asses his usefulness in practice. 
Indeed, the fact that the model has the correct sharp-interface limit
does not guarantee its practical usefulness for several reasons. On the
one hand, the stability of both the bulk phases and the kink profile 
must be assured, since this might not be the case in general. On the other
hand, a direct empirical test is necessary to determine quantitatively 
how close a finite 
$\epsilon$ situation is to the sharp-interface limit. 
This means finding a set of explicit quantitative criteria to choose 
all the nonphysical parameters in order to ensure a desired accuracy. 
Finally, it
is interesting to evaluate the capability of that model  at providing
quantitative results
with reasonable computing efforts in actual simulations. 

The rest of the 
paper is organized as follows: In Sec. \ref{seceqs} we present the
phase-field equations and the boundary conditions, discretization and parameters
used. Some 
preliminary tests on phase conservation are carried out in 
Sec. \ref{secdroplets}.
In Sec. \ref{secrd} we show how
to choose these computational parameters in the linear regime and
how to obtain the growth rates in the numerical linear
dispersion
relation, which we compare satisfactorily
with the sharp interface limit and with a 
thin-interface model. Convergence with these parameters is also 
tested. Sec. \ref{secst} is devoted to the steady state. 
Saffman--Taylor fingers are obtained from single-mode initial 
conditions in the non-linear regime, and their velocities and 
widths are compared with previous results. 
Multifinger 
configurations arising from a random initial condition in the 
linear regime are obtained in Sec. \ref{secmf}, which are morphologically 
and dynamically consistent with existing evidence both 
from experiments \cite{maher} and simulations \cite{jaume}.
Finally, we discuss the applicability of the model and its possible
future extensions in Sec. \ref{secdis}. 

\section{Implementation of the model}
\label{seceqs}
We will consider a rectangular Hele--Shaw cell of width $W$ ($x$-direction) and
gap $b$
($z$-direction), containing two fluids 
with distinct viscosities
($\mu_1$, $\mu_2$) and densities ($\rho_1$, $\rho_2$). 
Both fluids are separated by an interface with surface tension $\sigma$,
and move under an effective
gravity $g_{eff}$ 
(negative $y$-direction)
and an injection velocity $V_\infty$ (positive $y$-direction).
It is convenient to introduce the stream function $\psi$ 
as the harmonic conjugate, which can be defined by $u_x=\partial_y \psi$,
$ u_y=-\partial_x \psi$, where $u_x$, $u_y$ are the
$x,y$ components of the fluid velocity field $\vec u$. In terms of the 
stream function the governing equations in dimensionless form read
\begin{equation}
\label{eq:laplace}
\nabla^2\psi=0
\end{equation}
\begin{equation}
\label{eq:discontinuity}
\psi_r(0^+)-\psi_r(0^-)=-\gamma-c[\psi_r(0^+)+\psi_r(0^-)]
\end{equation}
\begin{equation}
\label{eq:continuity}
\psi_s(0^+)=\psi_s(0^-)=-v_n,
\end{equation}
where $r$ (resp. $s$) is a coordinate normal (tangential) to the interface,
$0^\pm$ 
means on the interface coming from either fluid, 
the subscripts
stand for partial derivatives except for $v_n(s)$, which
is the normal velocity of the
interface,
and 
\begin{equation}
\label{eq:gammasharp}
\frac{\gamma (s)}{2}\equiv B\kappa_s+\hat y \cdot \hat s,
\end{equation}
with
$\kappa (s)$
 the interface curvature.
The dynamics are controlled by the two dimensionless parameters
\begin{equation}
\label{eq:bandc}
B=\frac{b^2\sigma}{12W^2[V_\infty(\mu_1-\mu_2)+g_{eff} \frac{b^2}{12}(\rho_1- \rho_2)]},
\;\;\;\;\;\;\;\; c=\frac{\mu_1-\mu_2}{\mu_1+\mu_2}.
\end{equation}
Here, $c$ is the viscosity
contrast, that without loss of generality is taken as positive ($0\le c\le1$), 
and $B$ is a dimensionless surface tension,
measuring the ratio between the stablizing force of the capilarity
and the destabilizing 
driving force induced by injection and gravity.
We will restrict ourselves to the unstable case, i.e. positive values of $B$.
The equations are then written in the frame moving with the fluid
at infinity (or, equivalently, with the mean interface), and taking $W$ as the 
length unit
and $U_*\equiv cV_\infty+ g_{eff}\frac{b^2(\rho1-\rho2)}{12(\mu_1+\mu_2)}$ the velocity unit
(see \cite{aref1}).

The corresponding phase-field model which we have used was proposed in
\cite{teoric},
with $\theta$ the phase field:

\begin{equation}
\label{eq:sf}
\tilde{\epsilon} \frac{\partial\psi}{\partial t}=\nabla^2\psi+c\vec \nabla \cdot
(\theta \vec \nabla \psi)+\frac{1}{\epsilon} \frac{1}{2\sqrt 2}
\gamma(\theta )(1-\theta^2)
\end{equation}

\begin{equation}
\label{eq:pf}
\epsilon^2 \frac{\partial \theta}{\partial t}=f(\theta)+\epsilon^2\nabla^2\theta
+\epsilon^2 \kappa(\theta ) |\vec \nabla \theta |+\epsilon^2 
\hat z \cdot
(\vec \nabla \psi \times \vec \nabla \theta)
\end{equation}
where $f(\theta )\equiv \theta (1-\theta^2)$, and 
$\frac{\gamma(\theta)}{2}\equiv \hat s(\theta)\cdot(B\vec\nabla
\kappa(\theta)
+\hat y)$, 
$\kappa(\theta)\equiv -\vec\nabla \cdot \hat r(\theta)$,
with 
$\hat r(\theta)\equiv \frac{\vec\nabla \theta}{|\vec\nabla \theta|}$
and $\hat s(\theta)\equiv \hat r(\theta) \times \hat
z$, together with the boundary condition

\begin{equation}
\label{eq:bc}
\theta(y\rightarrow \pm \infty)=\pm 1,
\end{equation}
where $\theta=+1 (-1)$ corresponds to fluid 1 (2).

$\tilde{\epsilon}$ can be regarded as a 
diffusion time of the stream function over a
characteristic length of wavenumber $k$, 
$\frac{\tilde{\epsilon}}{(1\pm c)k^2}$ (which
must be chosen much smaller than the characteristic inverse 
growth rate of the 
corresponding mode of the interface
$|\omega|^{-1}$). On the other hand $\epsilon$ is basically the 
interface thickness (which
is required to be much smaller than the 
length scale
$|k|^{-1}$).
$\epsilon^2$ also stands for the relaxation time
of the phase field towards the kink profile, 
which in turn must be kept well below the inverse growth rate 
$|\omega|^{-1}$ (see Ref. \cite{teoric}. Sec. IV). These inequalities control
the convergence to the Hele--Shaw dynamics,
and, for the linear regime, the deviation
from it can be computed analytically
from them (see Ref. \cite{teoric}. Sec. IV),
thus providing {\it a priori} criteria for the choice of
$\tilde{\epsilon}$ and $\epsilon$ to
obtain a prescribed accuracy.

To be consistent with the frame change and adimensionalization inherent
to the parameters $B$ and $c$ previously defined we use a 
{ cell of width 1}, with periodic boundary conditions in the $x$-direction
(which include rigid walls as a particular case) {and no flux} (constant stream
function, e.g. $\psi=0$) { in the $y$-direction}. 
Whenever some symmetry of an initial condition exists and is preserved by 
the time evolution, we use a properly reduced integration domain 
(e.g. left-right symmetry for a single mode, or
up-down symmetry for an up-down symmetric initial condition with $c=0$).

Numerical integration of the above equations has been carried out on a grid 
of size $n_x \times n_y$ with equal spacing
$\Delta x=\Delta y=1/n_x$ in both x and y directions, 
using an explicit centered-space
algorithm.
$n_x$ is choosen so that
$\Delta x$ at most equals 
$\epsilon $ for the profile of the
fields across the interface to be properly
resolved. 
The $\epsilon $ value thus constrains $\Delta x$
and, in turn, the minimum size of the system
and the
maximum time step, so that the computing time required
to integrate a certain physical dynamics goes
As $\epsilon^{-4}$. 
Then $n_y$ is chosen so that there is always
a distance of at least one of the largest  present
wavelengths between the interface and 
the $\psi=0$ boundary condition at the end of
the channel. An increase in this distance does
not seem to affect appreciably the interface
evolution. This is what expected in the linear regime,
where the stream function decays exponentially
with the present wavelengths, but is an
{\it a posteriori} observation for the
Saffman--Taylor fingers.



As for the initial condition, in general one should set the phase field to 
$\theta=\tanh\frac{r}{\sqrt 2\epsilon}$, with $r$ being the signed distance to 
the desired interface, which is the model solution at least up to first order
in 
$\epsilon$ (see Ref. \cite{teoric}). Then the corresponding stream function
should be close to the stationary solution of Eq. (\ref{eq:sf}) for that 
phase field. This can be obtained either by solving the stationary version of
Eq. (\ref{eq:sf}) (a Poisson equation) or by letting the stream function in
Eq. (\ref{eq:sf}) relax while keeping the phase field frozen.
However in the present paper, since we will consider perturbed planar interfaces
as initial situations, we will make use of the fields predicted 
by the linear theory:
$\theta=\tanh\frac{y}{\sqrt 2\epsilon}$,
where $y$ is signed vertical distance to the desired interface,
and the thin-interface result (see Ref. \cite{teoric})
\begin{equation}
\label{eq:psilineal}
\psi(x,y)=a_\pm Ae^{ikx-q_\pm |y|},
\end{equation}
with
\begin{equation}
\label{eq:a1}
a_\pm =\frac{i\omega_0}{k}(\frac{1}{\sqrt{1+\frac{\tilde{\epsilon}\omega}{k^2}}}
\mp c\epsilon|k|{\sqrt 2})+{\cal O}(c^2)+{\cal O}(\epsilon^2).
\end{equation}
and
\begin{equation}
\label{eq:q}
q_\pm=\left|k
\sqrt{1+\frac{\tilde{\epsilon}\omega}{k^2(1\pm c)}}\right|.
\end{equation}

\section{Phase conservation}
\label{secdroplets}

Before testing the model in physical situations a 
preliminary test is that of the conservation of the two fluid phases.
Due to incompressibility and immiscibility, the Hele--Shaw dynamics
preserve the area and phase purity of each domain. In contrast, model
A, starting point for the equation for the phase-field Eq. (\ref{eq:pf})
(see Ref. \cite{teoric}), does not even preserve the global balance between
phases. For reasonable values of the interface thickness 
$\epsilon$ these non-conserved dynamics
are given aproximately by Allen--Cahn law ($v_n\propto \kappa$). Thus, 
phase-conservation within the phase-field scheme depends on the accuracy
of both the 
$\epsilon^2 \kappa(\theta ) |\vec \nabla \theta |$ term 
in cancelling
out Allen--Cahn law and the coupling
$\hat z \cdot
(\vec \nabla \psi \times \vec \nabla \theta)$
in introducing the Hele--Shaw dynamics.

First, we test Allen--Cahn law cancellation by considering situations in
which the stream function drops from the equations.
For a circular droplet of one phase embedded in the other
in absence of gravity
Eq. (\ref{eq:pf}) reduces to $\epsilon^2\frac{\partial \theta}
{\partial t}=
f(\theta)+\epsilon^2\frac{d^2\theta}{dr^2}$ 
(because $\frac{\partial\theta}{\partial s}=0$),
where $r$ is the radial coordinate.
This is exactly model A in one dimension, and therefore Allen--Cahn law
cannot arise.
Starting with 
$\theta=\tanh\frac{r}{\sqrt 2\epsilon}$ as 
initial condition, there is some slow dynamics which soon gets stacked because
of the limited numerical resolution. From then on, the fluid phases are
completely conserved. In contrast, if the term cancelling out Allen--Cahn law
is removed from Eq. (\ref{eq:pf}), the droplet quickly collapses.

The same test can be carried out for a marginal mode in a nearly flat
 interface with identical results: 
If the Allen--Cahn law correction is removed, the mode quickly
flatens, whereas, if it is not, some slow dynamics soon gets
stacked within the numerical resolution. This corresponds to the
point $k'=1$ in Fig. 2 (see next section),
 whose measured growth rate $\omega'$ vanishes exactly.
Both tests suggest that the
$\epsilon^2 \kappa(\theta ) |\vec \nabla \theta |$ term cancels
out Allen--Cahn law even at higher orders in $\epsilon$ than
those computed in Ref. \cite{teoric}.

As for general situations in which the coupling term can also play
a role, significant violation of phase conservation has only been
observed for such large values of
$\epsilon$ that 
 $\epsilon\kappa>1$. 
We therefore conclude that the two fluid phases are conserved
for reasonable values of $\epsilon$, and that, in practice, this is not a
restrictive condition on $\epsilon$.

\section{Linear dispersion relation}
\label{secrd}
The first physically relevant situation in which we have tested the model is the
linear regime of a
perturbed planar interface. 
The linear dispersion relation has been 
computed for vanishing viscosity contrast 
($c=0$). 
Sharp-interface model predicts a linear growth that does not depend on $c$.
However, the phase field model should 
exhibit some dependence in the viscosity
contrast related to the finite-$\epsilon$
and -$\tilde{\epsilon}$ corrections 
(see Ref. \cite{teoric}. 
Sec. IV).

We use a single mode occupying the whole channel width (i.e., of 
wavelength 1 and wavevector $k=2\pi$) and then vary
the dimensonless surface tension $B$ in order 
to change the growth rate of that mode according to the Hele--Shaw
dispersion relation
\begin{equation}
\label{eq:rdhs}
\omega_0=|k|(1-Bk^2).
\end{equation}
This is physically completely equivalent to fixing the surface tension
and varying the wavevector of the mode, as can be seen through the
rescaling $k'=\sqrt B k, \omega'_0=\sqrt B \omega_0=|k'|(1-k'^2)$. However,
it is numerically more convenient, since it allows one to use the same
value of $\epsilon$ for all the modes, because $k$ is fixed.
Thus, according to the finite-$\epsilon$
and -$\tilde{\epsilon}$ dispersion relation
derived in Ref. \cite{teoric}
\begin{equation}
\label{eq:rdthin}
\omega=\omega_0(\frac{1}{\sqrt{1+\frac{\tilde{\epsilon}\omega}{k^2}}}
-\epsilon|k|{\sqrt 2}\frac{5}{6})+{\cal O}(c^2)+{\cal O}(\epsilon^2).
\end{equation}
We have used $\epsilon=0.01$, $\tilde{\epsilon}=0.1$ in order to keep
the deviation from the sharp-interface one, $\omega_0$, below a 10\%
(we have used its maximum, $\omega=2\pi$,
to compute the required value of $\tilde{\epsilon}$).
Finally, the initial amplitude of the mode
is chosen
to equal the mesh size $\Delta x$ ---which in turn is set to $\epsilon/2$ for
reasons explained below---, so that we stay for a while in the linear regime
for unstable 
modes, and yet we have enough numerical subgrid resolution for the stable
ones also for a while.

The time evolution of a relevant mode (one close to
the most unstable one), with $B=6.5\times 10^{-3}$, is shown in Fig. 1. The points
(+) have been obtained
by averaging the ratio of the interface height $y$ at a certain time $t$
and the initial one $y(0)$ over 1/4th
of the interface.
The slope of the resulting curve in a linear-log plot is the growth rate. Where 
this is constant, the growth is exponential and we are in the linear regime.
This corresponds to the linear fit plotted in Fig. 1., from which we obtain 
the growth rates shown in Fig. 2. Beyond this regime the points curve down because
of the 
non-linearities. The crossover corresponds to an amplitude of about a tenth of the
wavelength of the mode.

The behaviour of the stream function has been plotted
in a dotted line
for purposes of comparison.
According to the linear theory (see, e.g.
\cite{teoric}),
the absolute value of the stream function
at any point of the system
should also grow exponentially
with the same growth rate than the
interface, and this is indeed the case,
since the dotted line ---average 
over the whole bulk of the 
absolute value of the ratio between 
the stream function $\psi$
in a certain point
at a certain time $t$ and the initial one
at the same point $\psi(0)$---
is parallel
to the interface evolution in the linear
regime. The gap between the two straight
lines is due to the initial, very
short ---and therefore not visible
in the figure--- decay of the
stream function, which reflects the 
rounding on a scale of
${\cal O}(\epsilon)$ of the gradient
on the interface of the sharp-interface
result used
as initial condition.

In Fig. 2 we present the linear dispersion
relation thus obtained in the rescaled variables defined above.
The points (+) correspond to the growth rates measured as in Fig. 1,
for times ranging from 0.3 to 0.7, i.e., for roughly 
a decade in the amplitude. Their deviation from the sharp-interface
result of Eq. (\ref{eq:rdhs}) (solid line) keeps below the desired 10\% error 
and is fairly well { quantitatively}
predicted by the thin-interface dispersion relation of
Eq. (\ref{eq:rdthin}) 
(dotted line). This quantitative agreement between theory 
and numerics is quite remarkable
if we take into account that the thin-interface
model is based on an asymptotic expansion in $\epsilon$. 
This good agreement is indeed an indication that the value of
$\epsilon$ used is in the asymptotic regime of the sharp-interface limit,
as we will see more clearly in Fig. 3.

Although we in fact introduce the
thin-interface
growth rate Eq. (\ref{eq:rdthin}) into the initial condition for the 
stream function, 
the results do
not depend on this after a certain transient. If one sets more naively the
sharp-interface growth rate $\omega_0$ into the initial condition for the
stream function, one gets a rather long transient in which the slope of an
amplitude-time log-linear plot like that of Fig. 1 initially curves down (up)
for an unstable (stable) mode, i.e., the growth is not exponential, and we are
therefore not describing properly the Hele--Shaw problem. The slope
then relaxes towards the thin-interface one and the growth becomes exponential,
but non-linearities soon set in. In contrast, if the thin-interface growth
rate is set into the initial stream function, one gets constant slopes
from the very beginning. This allows one to measure the growth rate in a 
wider range of time and is indeed also an indication that the thin-interface
model
is valid.

The growth rate values shown in Fig. 2 could still be refined by further
decreasing $\epsilon$
and $\tilde{\epsilon}$. This is not only a theoretical possibility, but can
also be done in practice as we show in Fig. 3, although the computation
time increases as explained above. Here, we study the
convergence of the 
growth rate $\omega$ (y-axis)
for the maximum of $\omega'(k')$ ($B=8.443\times 10^{-3}$)
to
the Hele--Shaw result $\omega_0=4.188$ (left upper corner)
as we decrease $\epsilon$ (x-axis) and $\tilde{\epsilon}$
(various symbols).
The empty symbols have been obtained with $\Delta x=\epsilon$,
whereas the filled ones correspond to $\Delta x=\epsilon/2$.
The growth rates obtained with $\Delta x=\epsilon$ are always below
the ones for $\Delta x=\epsilon/2$, probably because of the
stabilizing effect of the mesh size.
If $\Delta x$ is
further decreased the differences with the values computed with
$\Delta x=\epsilon/2$ are tiny, whereas the gap between the $ \Delta x=\epsilon$
and the $\Delta x=\epsilon/2$ points is pretty large (clearly more than
the differences between distinct
symbols ---distinct $\tilde{\epsilon}$ values--- or adjacent values
of $\epsilon$). This means that the discretization has practically converged
to the continuum model for $\Delta x=\epsilon/2$, but not for
$\Delta x=\epsilon$. That is the reason why we have used $ \Delta x=\epsilon/2$
in Figs. 1 and 2. 

Moreover, the $\Delta x=\epsilon/2$ points should
Be described by the thin-interface model of Eq. (\ref{eq:rdthin}), 
and this is indeed the case for small enough values of $\epsilon$.
To visually see this we have plotted the thin-interface prediction
for $\tilde{\epsilon}=0$ (dashed line), which is, of course, 
a straight line in $\epsilon$. Each set of points with the same
$\tilde{\epsilon}$ value clearly tend to allign parallel to this
line as $\epsilon$ decreases ---as Eq. (\ref{eq:rdthin}) predicts---,
whereas they curve up and even cross the line for large values of
$\epsilon$, for which we are beyond the asymptotic regime of
validity of Eq. (\ref{eq:rdthin}), apparently ending near 
$\epsilon=0.01$. This makes $\epsilon=0.01$ very suitable for
simulations, and confirms it to be within 
the asymptotic regime as we pointed out above.
Note as well how the growth rate increases with decreasing values
of $\tilde{\epsilon}$ within the same value of $\epsilon$ 
(vertical columns of points), and how it approaches the dashed line
in good agreement with the values predicted by Eq. (\ref{eq:rdthin})
for values of $\epsilon$ within the asymptotic regime.

\section{Saffman--Taylor fingers}

\label{secst}
Once the model has been tested in the linear regime,
it seems mandatory to check a highly non-linear situation such
as the steady state, for which many analytical and numerical results are
available. 

We start with a single mode occupying the whole channel and with
an amplitude equal to the channel width, i.e.,
clearly in the non-linear regime. Obviously the initial
condition predicted
by the linear theory is not accurate, but this just introduces
a transient during which the dynamics is not the Hele--Shaw one.
By the time the forming finger is close to the steady shape, all transients must
have decayed.
Moreover, starting with a single mode we
not only check that the model exhibits a
finger solution, which we could do much faster by starting with
that solution, but 
that the model is robust to inaccurate initial conditions for the bulk, 
and also that single fingers have a reasonable
basin of attraction, which may be a rather nontrivial point for small $c$
\cite{jaume}.

Here we do not have a theoretical prediction for the deviations from
the Hele--Shaw steady state due to finite-$\epsilon$ and -$\tilde{\epsilon}$
effects. However, we know that the inequalities of Sec. \ref{seceqs}
still control
the convergence. The differences are that $\omega$ now stands for the
steady advance velocity (which should be in the range 0.7-0.8, not far 
from the maximum growth rate in the linear regime $\omega=2\pi$), and
that all unstable modes ---and not only the one set by the initial
condition--- are present in the steady state, so that $k$ must be choosen 
to be the less favourable in each inequality. This continues to be $k=2\pi$
for the condition on $\tilde{\epsilon}$, but becomes $k=1/\sqrt{3B}$, the
most unstable mode, for the one on $\epsilon$. We thus
progressively decrease the values of 
$\epsilon$ and $\tilde{\epsilon}$ and look for convergence (in the
finger widths) within the 
computational resources.

The finest comparable simulation runs are
shown in Figs.
4.a-d. For $B=10^{-2}$ (Figs. 4.a-c) we have used $\epsilon=0.02$, twice
less accurate than the one used in the linear regime (Figs. 1-2), whereas
for $B=10^{-3}$ (Fig. 4.d) $\epsilon=0.01$, three times less accurate.
As for $\tilde{\epsilon}$, we have used $\tilde{\epsilon}=0.2$ for $c=0$
(Figs. 4.a and 4.d), $\tilde{\epsilon}=0.1$ for $c=0.5$ (Fig. 4.b), and
$\tilde{\epsilon}=0.02$ for $c=0.9$ (Fig. 4.c), all of them twice less 
accurate than the one in the linear regime. However, convergence
in $\tilde{\epsilon}$ seems to
be achieved, at least in the finger widths. We cannot assure it for  
$\epsilon$ because the finger tip splits for $B=10^{-3}$ 
---probably because of numerical
noise--- if $\epsilon$ is further lowered.
In all cases we have used $\Delta x=\epsilon/2$ like in the linear regime tests,
although we have found that already 
for $\Delta x=\epsilon$, the differences in widths
are at most within a 1\%.

Figs. 4.a-d show single fingers that are formed and
relax to a certain width to 
just advance with a steady velocity. This can be measured plotting the 
tip position against time, finding excellent linearity from $t=1.15$
to at least
$t=3.3$ (for the fastest finger). Four successive interfaces at
constant time intervals are shown
covering this whole regime. Except for the $c=0.9$ case of Fig. 4.c, 
the finger has not completely relaxed to its final width
for the first plotted interface, which
means that we are not in the steady state yet. From the second to 
the last shown interfaces, no observable 
change in width, tip velocity or shape is seen  
anymore. Taking into account that Figs. 4.a-d show the whole channel
---although simulations were carried out using half ($c\ne 0$)
or a quarter ($c=0$) of it---
it is clear from Fig. 4.d 
that the end of the channel can be set even closer to the tip
of the finger than a channel width (i.e., closer than what the linear
regime would have suggested).
Regarding to the values of this steady velocity, they are always a 10-20\%
below what predicted for an (infinite) Saffman--Taylor finger
of the same width without surface tension,
although velocity does increase with decreasing
finger width as can be seen comparing Figs 4.a and 4.d.
The origin of this quantitative discrepancy must lie necessarily 
on  finite-$\epsilon$ and -$\tilde{\epsilon}$
effects, most likely related to the non-laplacian character introduced 
by $\tilde{\epsilon}$, which produces a finite diffusion length.

Note that $c=0$ fingers (Figs. 4.a and 4.d) are up-down symmetric (also
if one uses the whole cell as integration domain), whereas the $c\ne 0$ ones
are not, but the less viscous fluid (down, negative $y$ values)
propagates into the more
viscous one (up, positive $y$ values)
in the shape of a Saffman--Taylor finger,
whereas the more
viscous propagates into the less one in the shape of a drop.
If Fig. 4.a is compared to 4.b and 4.c (all with 
$B=10^{-2}$ but increasing viscosity contrast $c$), one sees that 
the higher the viscosity contrast, the longer the Saffman--Taylor finger
and the shorter the drop part (the initial condition 
was centered about y=0).
All such dependences on $c$ coincide with Tryggvason and Aref's observations
from their vortex sheet numerical scheme \cite{aref2}.

As for the finger widths, all above 1/2 of that of the channel,
they decrease with decreasing dimensionless surface
tension $B$ and increasing viscosity contrast $c$. This $c$-dependence
can also be seen in Fig. 1 of Ref. \cite{aref2}. Quantitative comparison
with McLean--Saffman numerical solvability theory results for $c=1$
is only feasible to some extent if $c=1$ widths are extrapolated from
the $c\ne 1$ ones obtained, since our model cannot treat the limiting 
case $c = 1$. By doing this for $B=10^{-2}$ ($\lambda=0.63$ for $c=0$, 
$\lambda=0.61$ for $c=0.5$, and $\lambda=0.60$ for $c=0.9$, Figs. 4.a-c),
one obtains $\lambda\simeq 0.60$ for $c=1$, a 6\% below the 
McLean--Saffman value $\lambda=0.637$. For $B=10^{-3}$, $c>0$ fingers
would become very computationally demanding, but if the same gap 
between $c=0$ and $c=1$ widths as for $B=10^{-2}$ is assumed, this would
yield $\lambda\sim 0.56$ for $B=10^{-3}$, $c=1$, within a 5\% distance
of the McLean--Saffman value $\lambda=0.528$. Of course this latter 
extrapolation is not rigorous, since the differences in width due to 
the viscosity contrast are likely to diminish with decreasing $B$, so
that they vanish for $B\rightarrow 0$ and $\lambda=1/2$ 
is selected for all $c$, as is generally believed to be the case.
In any case, the deviations from the McLean--Saffman results found 
(within a 5-6\%) are surprisingly lower than what the used values of 
$\epsilon$ and $\tilde{\epsilon}$ would have suggested taking into
account the observed deviations in the linear regime.

\section{Multi-finger dynamics}
\label{secmf}

We have finally tested the model in the non-linear dynamics appearing between the
precedent situations, i.e. after the departure of the linear regime and before
reaching the steady state.
Here we use a somewhat experimentally realistic initial
condition consisting of a superposition of sinusoidal modes
with random, uniformly distributed
amplitudes between -0.005 and +0.005 for each wavelength
$\lambda=1,\frac{1}{2},\frac{1}{3}...\frac{1}{7}$ ---i.e., in the
linear regime and random phases. For $B=10^{-3}$ the most amplified of these 
wavelengths will be $\lambda=\frac{1}{3}$, so that we expect
3 unequal fingers to appear and there is a chance for mode
interaction and competition
to set in. Wavelengths below 
$\lambda=0.161$ are stable and will decay. We include some of them anyway.
Then all modes are added up to
find the interface position. The stream function predicted by the
linear theory is also obtained by adding up each mode's
stream function, but all with their peaks centered at the same
final interface position, to avoid the formation of more than one peak of
the stream function across the interface.

Since harmonics of the channel width are present, we have to
refine the $\epsilon$ used in the linear regime. We use 
$\epsilon=0.00625$, with
$\Delta x=\epsilon$ to save computation time. The value of 
$\tilde{\epsilon}$ is quite crude ($\tilde{\epsilon}=0.5$
for the equal viscosities case $c=0$, Fig. 5.a), especially for
the high viscosity contrast ($c=0.8$) run (Fig. 5.b)
($\tilde{\epsilon}=0.2$). 

The results are shown in Figs. 5.a,b for $0.15<t<1.25$ at constant
time intervals 0.1 (dots). The latest interface is emphasized in
bold and the solid line corresponds to the initial
condition for $t=0$.
As we can see, the initial condition happens to have six maxima.
Rather quickly, only three of them are left as predicted by linear stability, 
even before entering the non-linear regime, in which these maxima
elongate into well developed fingers. For vanishing viscosity
contrast ($c=0$, Fig. 5.a), there is no apparent competition,
in agreement with experimental \cite{maher} and numerical
 \cite{aref1,jaume} evidence. Longer and shorter 
fingers all advance. Shorter fingers might not advance so quickly,
but they expand to the sides, so they clearly keep growing.
In contrast, starting with the same initial condition that for 
$c=0$, the $c=0.8$ run (Fig. 5.b) shows competition between fingers
of the less viscous fluid advancing into the more viscous one
---and not the other way round---, as is known to happen in
the Hele--Shaw problem. The shorter finger now also expands 
laterally, but it soon begins to move backwards as a whole.
Longer times may lead to the pinch-off of droplets in both cases.

\section{Concluding remarks}
\label{secdis}
We have numerically integrated the phase-field model proposed in Ref.
\cite{teoric} for the Hele-Shaw problem in the unstable configuration. 
Simulations have been carried out in several situations: a circular droplet (to
test conservation of each fluid), linear regime of a perturbed planar interface,
steady state, and non-linear regime in between.
The performed tests guarantee that the model can be used in
practice to reproduce the Hele--Shaw dynamics. 
In the model two small parameters, $\epsilon$ and $\tilde{\epsilon}$, whose zero
limit correspond to the sharp interface limit,
can be independently chosen, being the dynamics well described by the
thin-interface
model derived in \cite{teoric}. In general, the conditions on $\epsilon$,
$\tilde{\epsilon}$
allow one to control the accuracy of the simulations. 


The basic criteria to control the closeness to the sharp interface limit 
are $\epsilon\kappa<<1$, 
$\frac{\tilde{\epsilon}\omega}{(1\pm c)k^2}<<1$. More precisely, in the linear
regime we find 
that both numerical simulations and the thin-interface model (with almost
identical results) are accurate  
with an error below 10\% if one satisfies the conditions
$\epsilon k \leq 0.06$, $\frac{\tilde{\epsilon}\omega}{(1\pm c)k^2}\leq 0.016$.
In this situation a run such as that leading to Fig. 1 typically took about 3
C.P.U. hours on a Pentium Pro at 200 MHz. On the other hand, in general the
potentially most
costly situation is the multifinger dynamics. 
Here, Fig. 5.a took about 70 C.P.U hours in the same processor.
 The method
could be made more efficient by using a semi-implicit scheme,
 an adaptative mesh, or 
(possibly) by
cancelling out the corrections to the sharp-interface equations
remaining (i.e., other than the Allen--Cahn law)
 in the thin-interface model. For high viscosity contrasts,
$c\sim 1$, a distinct model could possibly be more efficient.

Extensions to less studied, related problems are also enlighted, 
including the study of noise and spatial disorder (for instance to 
simulate porous
media), 
the study of liquid crystals and other complex fluids, and problems
with a physically diffuse interface, such as thermal plumes\cite{benamar} 
or motion of the salt/fresh water interface in coastal aquifers.

\section*{Acknowledgements}
We acknowledge financial support from the Direcci\'on
General de Ense\~{n}anza Superior (Spain), Projects No. PB96-1001-C02-02,
PB96-0378-C02-01 and PB96-0241-C02-02,
and the European Commission Project No. ERB FMRX-CT96-0085.
 R.F. also aknowledges
 a grant from the Comissionat per a Universitats i Recerca
 (Generalitat de Catalunya).

\newpage
\section*{figure captions}

\paragraph*{Fig. 1}
Measure of the interface growth rate for $B=6.5\times 10^{-3}$.
Crosses (+) correspond to the average over 1/4th of the 
interface of $y(t)/y(0)$ ($y\equiv$interface height) and the dotted
line to the average over 1/4th of the bulk of
$|\psi(t)/\psi(0)|$. The growth rate is the slope of the linear
fit (solid line).

\paragraph*{Fig. 2}
Linear dispersion relation in appropiate variables
(see explanation in text). Crosses (+) represent the growth
rates
obtained as in Fig. 1. The Hele--Shaw result (solid line)
and the thin-interface prediction of Eq. (\ref{eq:rdthin}) (dotted
line) are shown for comparison.

\paragraph*{Fig. 3}
Convergence to the Hele--Shaw value for the maximum of
the linear dispersion relation curve, $B=8.44\times 10^{-3},
\;\;\omega_0=4.18$, as $\epsilon$ ($x$-axis) and 
$\tilde{\epsilon}$ (1 for circles, 0.25 for triangles and
0.1 for squares) are decreased. The dashed line corresponds to
the thin-interface prediction of Eq. (\ref{eq:rdthin}) for 
$\tilde{\epsilon}=0$. Empty symbols have been obtained with
$\Delta x=\epsilon$, whereas filled ones used $\Delta x=\epsilon/2$.

\paragraph*{Fig. 4}
Saffman-Taylor finger advance: 4 successive interfaces
at constant time intervals for the regime in which the tip
velocity stays constant, $1.15<t<3.3$.
(a-c) $B=10^{-2},\;\;\epsilon=0.02$:
(a) $c=0,\;\;\tilde{\epsilon}=0.2\;\;(\lambda=0.63)$;
(b) $c=0.5,\;\;\tilde{\epsilon}=0.1\;\;(\lambda=0.61)$;
(c) $c=0.9,\;\;\tilde{\epsilon}=0.02\;\;(\lambda=0.60)$.
(d) $B=10^{-3},\;\;\epsilon=0.01,
\;\;c=0,\;\;\tilde{\epsilon}=0.2\;\;(\lambda=0.59)$.

\paragraph*{Fig. 5}
Time evolution for $B=10^{-3},\;\;\Delta x=\epsilon=0.00625$
of an initial condition (solid line)
combining $\lambda=1,\frac{1}{2}..\frac{1}{7}$
with random, uniformly distributed
amplitudes between -0.005 and +0.005.
Interfaces at time intervals 0.1 are shown in
dotted lines starting at $t=0.15$. 
The latest interface shown ($t=1.25$) is represented in bold. 
(a) lack of competition for $c=0$ ($\tilde{\epsilon}=0.5$);
(b) competition for $c=0.8$ ($\tilde{\epsilon}=0.2$).

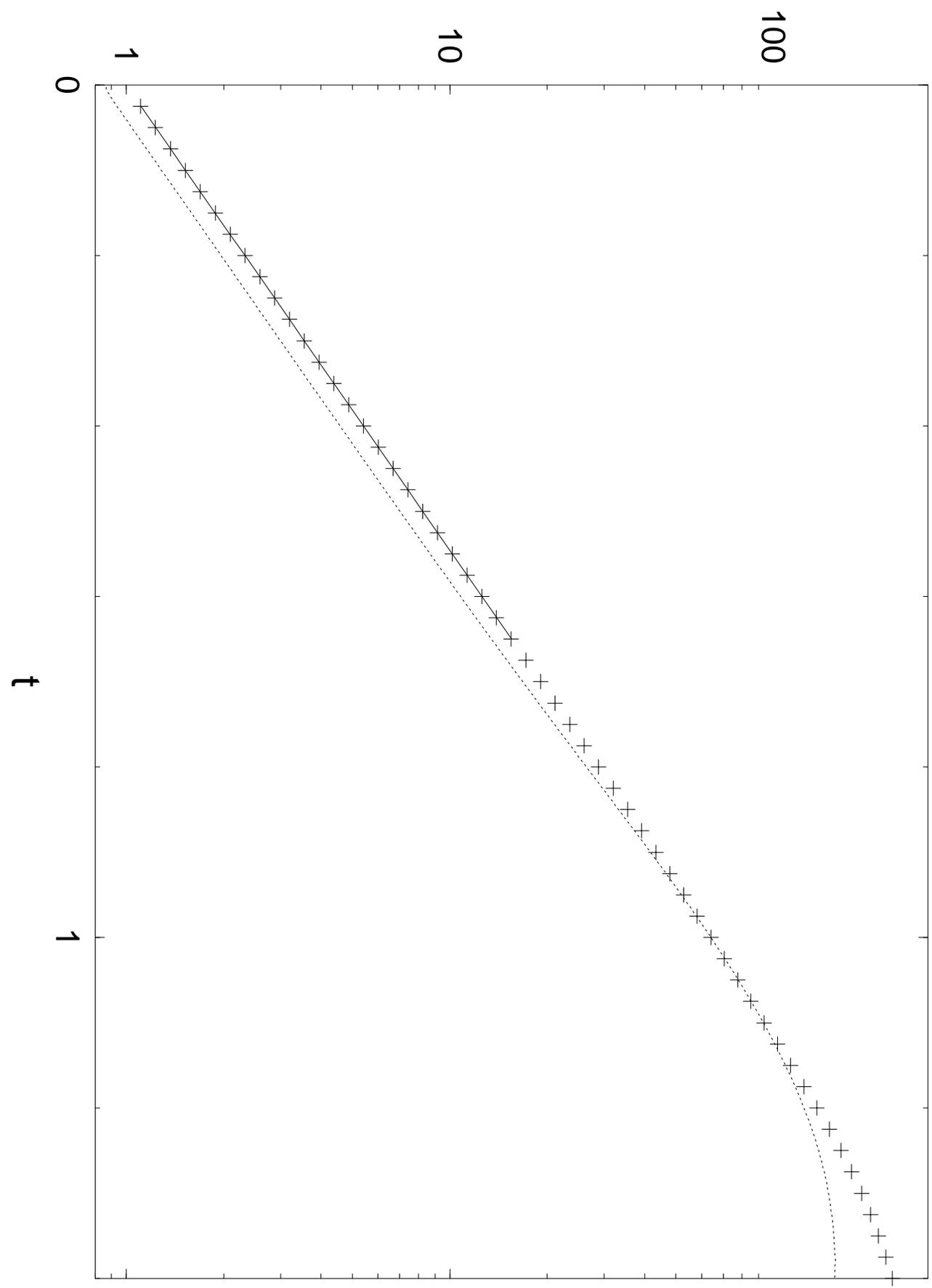
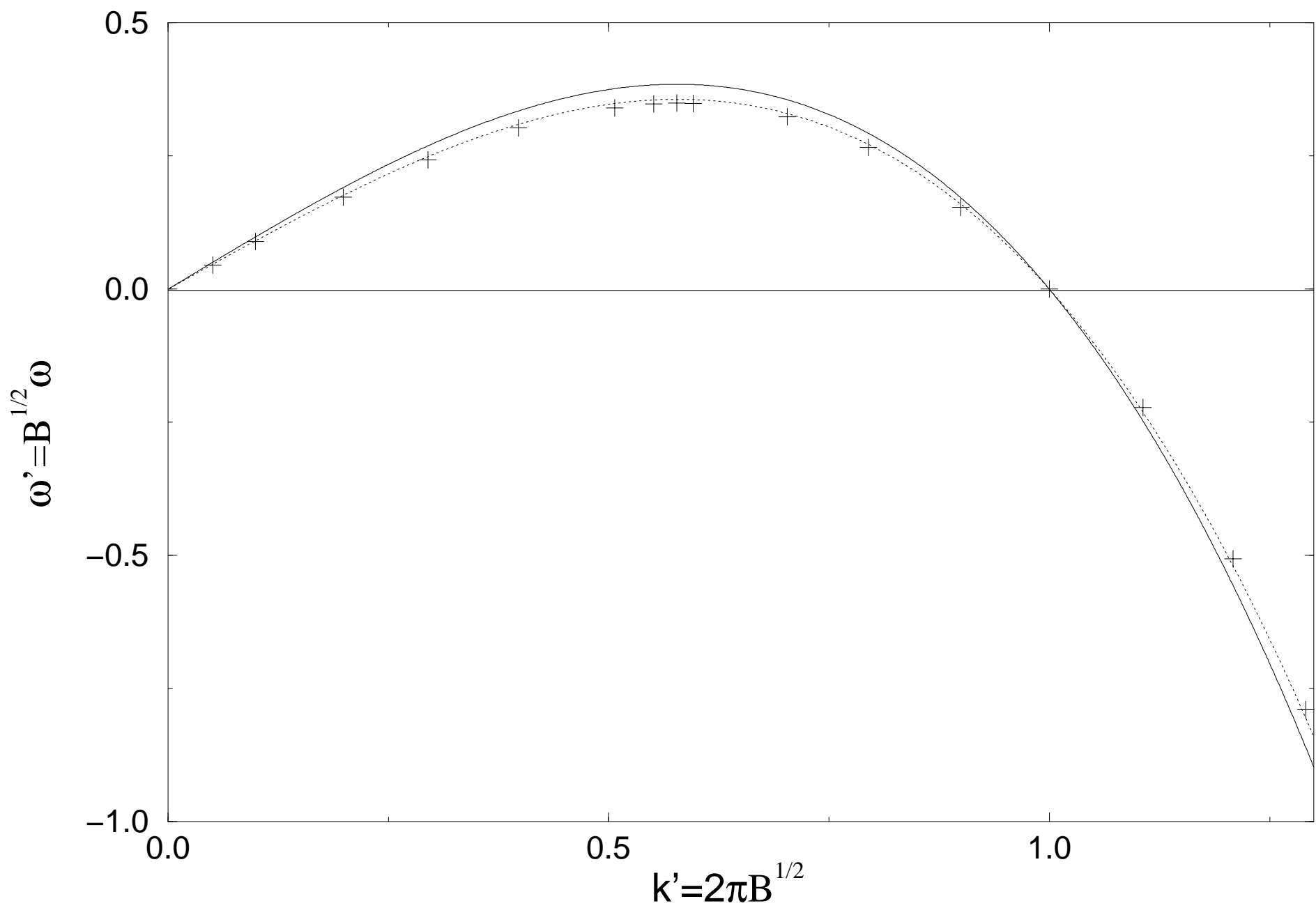
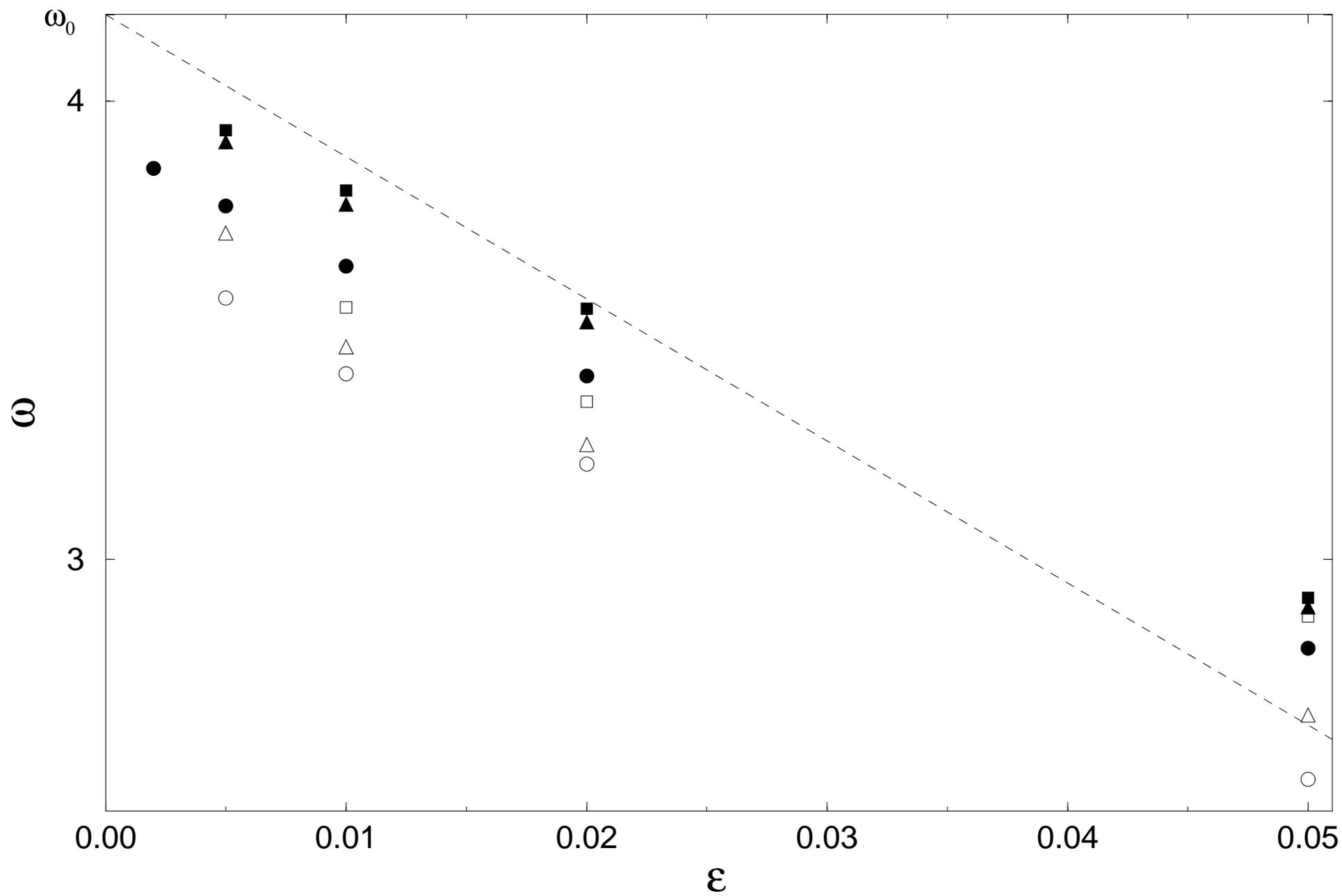
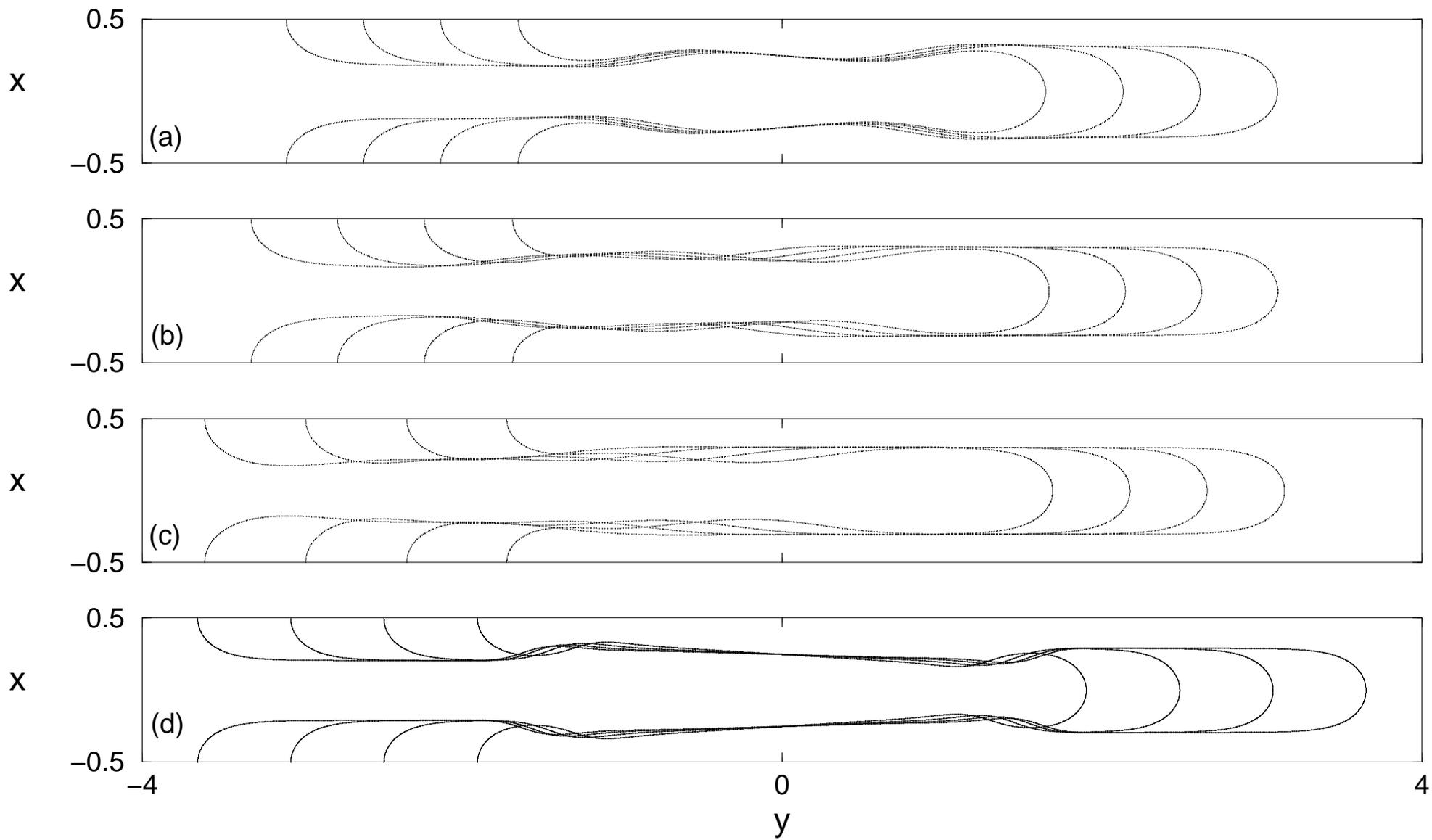
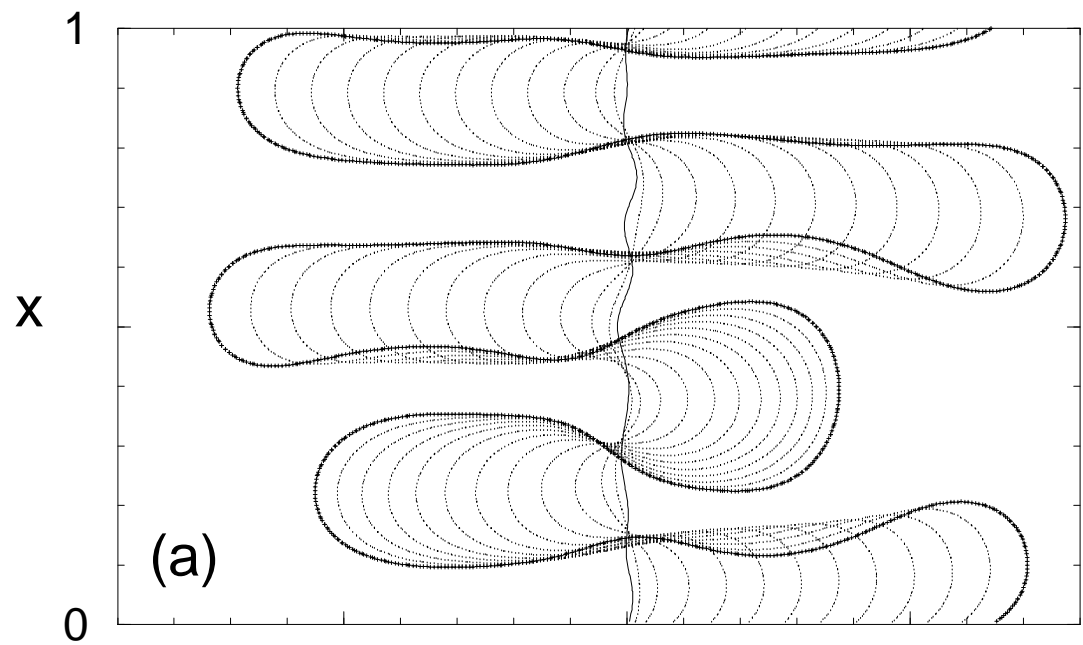
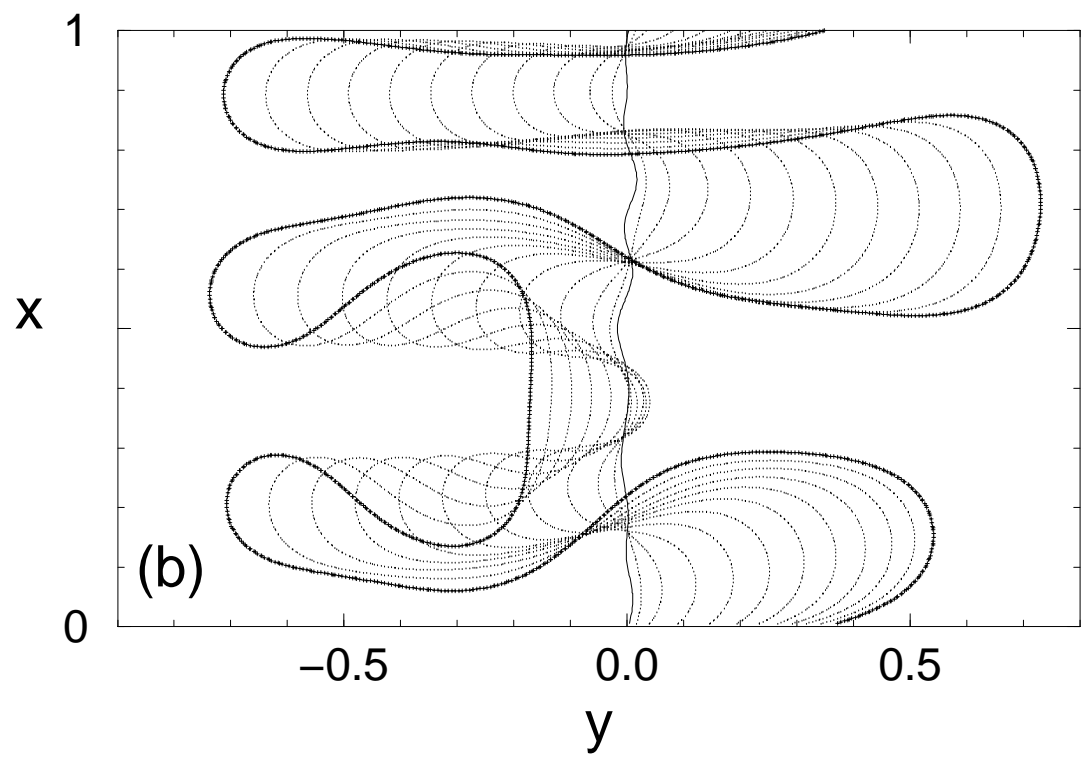
\end{document}